\documentstyle[12pt, amsmath, amsfonts]{article}

\newcommand{\set}[1]{{\mathbb{#1}}}

\newcommand{\bepsilon}{\mbox{\boldmath $\epsilon$}}
\newcommand{\br}{\mbox{\boldmath $r$}}
\newcommand{\bv}{\mbox{\boldmath $v$}}

\newcommand{\Jf}{{}^J\!f}
\newcommand{\ttS}{{\tt S}}
\newcommand{\ttU}{{\tt U}}
\newcommand{\ttV}{{\tt V}}

\newcommand{\ttu}{{\tt u}}
\newcommand{\ttv}{{\tt v}}

\newcommand{\oo}[1]{\overset{\circ}{#1}}
\newcommand{\ooo}[1]{\overset{\circ\circ}{#1}}

\begin{document}

\title{ \vspace{-1.8cm}
{Occam's Razor as a formal basis for a physical theory}}

\author{Andrei N. Soklakov\footnote{e-mail: a.soklakov@rhul.ac.uk}\\
\\
{\it Department of Mathematics}\\ 
{\it Royal Holloway, University of London}\\
{\it Egham, Surrey TW20 0EX, United Kingdom}}

\date{26 September 2001}

\maketitle

\begin{abstract}
  \vspace{-9mm}
  We introduce the principle of Occam's Razor in a form which
   can be used as a basis for economical formulations of physics.
   This allows us to explain the general structure of the Lagrangian
   for a composite physical system, as well as some other
   artificial postulates behind
   the variational formulations of physical laws. As an
   example, we derive Hamilton's principle of stationary action together
   with the Lagrangians for the cases of Newtonian mechanics,
   relativistic mechanics and a relativistic particle in an external
   gravitational field.

  PACS:
   45.05.+x, 
   45.20.-d, 
   45.50.-j,  
   89.70.+c  
   
\end{abstract}

\section{Introduction}

The standard derivation of the laws of motion from the principle
of least action is a cornerstone of all major physical
theories. However, this derivation is based on a number of postulates
which are too unnatural to be considered as axioms. For instance,
why at the fundamental level does the Lagrangian $L$ of a composite
system  always have the form $L=L^1+L^2-V$, where $L^1$ and $L^2$
are the Lagrangians of free subsystems and $V$ accounts for the
interaction? Indeed, what meaning can we assign to the difference
between kinetic and potential energy, which is used as a template for
many physical Lagrangians? Furthermore,
why is the action defined as an integral of $L$, and why do we
obtain correct classical equations by minimizing this integral? 
These questions are equally important in almost
every theory which uses Lagrangian formulation. We may therefore
hope that the answers to these questions can be useful in finding the
ultimate physical theory.

According to Occam's Razor, simple theories are more
economical and are usually better suited for making predictions.
Indeed, all fundamental laws of physics are surprisingly simple in form.
In this paper we introduce Occam's Razor in a form of a physical
principle that we call the {\it simplicity principle} ({\it SP}).
Using the {\it SP} we answer the above questions: we 
explain the structure of the Lagrangian of a
composite physical system together with the other
postulates behind the Hamilton's principle of stationary action.
In this sense we derive the Hamilton's principle of stationary action.

As a first
step, we introduce the standard notion of the state space of a
mechanical system and derive Newton's second law from just two extra
postulates, namely the {\it SP} and the Galilean relativity principle.
The purpose of this derivation is to demonstrate our approach using
the particularly well known case of Newtonian mechanics.
The important contribution from our approach is that the
mathematical structure of the {\it SP} alone implies that all
fundamental interactions can be accounted for
by adding an extra term to the Lagrangian. This result is {\it independent}
of the particular theory. In other words, different theories correspond
to different Lagrangians for free elementary systems, whereas
the {\it SP} tells us how to introduce interactions between them.
This means, for instance, that in applying our theory for
the relativistic case it is enough to consider one particle cases,
such as a free
relativistic particle and a relativistic particle in an external
gravitational field. The rest of the arguments follow simply by
replacing Galilean relativity with Einstein's principle of
relativity and his principle of equivalence. \\

\section{Dynamical laws and the Simplicity Principle } \label{sec:SP}

The task of theoretical physics is to find algorithms that can
correctly reproduce or predict experimental data. However, not every
such algorithm can be considered satisfactory, as real understanding
implies that a minimal set of simple axioms is found and all
experimental results can be reproduced as a consequence of these
axioms.  The axioms should introduce a {\it state model} able to
describe the system instantaneously, and {\it the
  dynamical laws} that describe any physical changes of the system's
state.  
Since the set of
evolution histories is incomparably more numerous than the set of system
states, the complexity of the system dynamics given the system state
model can be very large.  In this context it is not trivial
that all fundamental laws of physics should be simple in form, yet
this is true for {\it all known} fundamental laws, even ones as diverse as
gravitation and quantum mechanics.  In this paper we propose to use
this fact as a common ground for the laws of physics.

Given a physical system, consider the set $\set{S}=\{\xi\}$ of {\it
  all} states in which the system can be prepared or experimentally
found.  We thereby require that each $\xi\in\set{S}$ contains a {\it
  complete} description of a system state.  This means that there
can be no hidden information (such as the preparation history) that
would distinguish otherwise identical system states. A function $f:
\set{S}\to\set{S}\times\set{S}\times\set{S}\times\cdots$ is called a
dynamical law if, for any initial state $\xi_0\in\set{S}$, the value
$f(\xi_0)$ is an ordered sequence $\{\xi_1,\xi_2,\dots\}$,
$\xi_k\in\set{S}$.  Physically, $f(\xi_0)$ defines a trajectory in
$\set{S}$ associated with the initial state $\xi_0$.

By the definition of $\set{S}$, we assign the same physical meaning to a
system state regardless of its preparation history.  In other words,
if the system passes through an intermediate state $\xi_i$ then the
predicted evolution following $\xi_i$ should not depend on how the
system reached $\xi_i$. For each dynamical law $f$ that
satisfies this requirement we can find a function $g$ such that for
every two consecutive points $\xi_i$ and $\xi_{i+1}$ in the trajectory
$f(\xi_0)$ we have $\xi_{i+1}=g(\xi_i)$. For further reference we will
call such laws Markovian. It is clear that without loss of generality
we can consider only Markovian laws. This is because the preparation
history can always be included in the description of the system state,
in which case the the evolution will be Markovian.

Not all dynamical laws describe the actual system evolution equally well.
Considering the set of all possible dynamics as a hypothesis space,
we may follow one of the standard approaches in the
formal induction theory~\cite{LiVitanyi}.
For instance, we can try to find a single law which, by some criterion,
is better than any other law.
Alternatively, we can try to use the individual predictions
of each possible law and formulate our final prediction by averaging
over all individual predictions using some ``prior'' probability
distribution. In this paper we
follow the strategy of singling out only one dynamical law and
leave the second, more general approach, for the
discussion of further research on quantization (section \ref{Discussion}).

In order to discriminate between different dynamical laws, we
postulate that the most economical set of axioms for a physical theory
includes the simplicity principle ({\it SP}): {\it 
      among all dynamical
      laws that are consistent with all the other axioms, the laws with
      the smallest descriptional complexity predominate the system's
      behavior}. 
The {\it SP} has philosophical and historical roots in
Occam's Razor, which was stated by Isaac Newton as Rule I for natural
philosophy in his famous {\it Principia}. Relatively recently, Occam's
Razor became a cornerstone of the modern theory of induction and
computational learning as introduced by Solomonoff \cite{Solomonoff64}
in 1964 (see also Ref.~\cite{VitanyiLi} for a thorough review
including more recent developments).
 
In contrast to computational learning, we do not analyze a collection
of raw experimental data. Using the examples of Newtonian
and relativistic mechanics, we demonstrate that only the most
general axioms (such as the Galilean relativity principle or Einstein's
relativity principles) are sufficient to complete the theory if
combined with the {\it SP}.  Physics enters our formalism both through the
definition of the ``state space'' $\set{S}$ of the system and through the
relativity principles; these are taken as experimental facts. The {\it
  SP} provides an inference tool for finding the simplest dynamical
theory consistent with these experimental facts.

The main weakness of this paper is a rather artificial
proof that there is no contradiction in using Kolmogorov complexity to
quantify the complexity of dynamical laws in the case of Newtonian
mechanics.  The requirement of Galilean relativity appears as a
constraint on the complexity of physical dynamics. Even though this
constraint is mathematically consistent and can be satisfied, it
appears to be rather artificial in the framework of algorithmic
information theory. This does not occur in the relativistic case which
naturally follows from an absolutely analogous yet technically simpler
analysis.  We suspect that the difficulties in the Newtonian case
arise from the special role of time, although a more
suitable measure of complexity of dynamical laws probably
can be proposed.

\section{Mathematical background and key ideas } \label{definitions}

To formulate the {\it SP} mathematically we need a measure of
complexity which can be assigned to individual objects (as we need to
discriminate between particular laws).  In 1963-1965, A.~N.~Kolmogorov
(see e.g. \cite{Kolmogorov65}) proposed to consider this problem in
the framework of the general theory of algorithms. Similar results
were obtained by R.~J.~Solomonoff~\cite{Solomonoff60,Solomonoff64},
and by G.~J.~Chaitin \cite{Chaitin69};  these three authors had
different motivations and worked independently from one
another~\cite{LiVitanyi}. Significant progress has been made to
improve the original definitions of complexity so as to increase the
range of applications. For the purpose of this paper
we will use the prefix version
of Kolmogorov complexity which was introduced by Levin \cite{Levin74},
Gacs \cite{Gacs74} and Chaitin \cite{Chaitin75}.

All the key properties of prefix complexity that are necessary
for our results are summarized by
Eq.~(\ref{TheFormula}). This is important because
Eq.~(\ref{TheFormula}) represents a very typical property of
information in both algorithmic and probabilistic approaches. This
property is often illustrated by Venn diagrams which make
Eq.~(\ref{TheFormula}) a very natural requirement for any
information-theoretic measure of complexity.  In physics, we often
deal with integrable, differentiable and even smooth functions.
Kolmogorov complexity can be interpolated by such functions only in
special cases, and under some severe restrictions on its arguments.
This fact makes it difficult to work with Kolmogorov complexity even
though Eq.~(\ref{TheFormula}) is all we really need at this stage. We
therefore acknowledge that by using an alternative measure of
complexity which obeys Eq.~(\ref{TheFormula}) we may considerably
simplify the arguments of this paper.  In the first reading, we
recommend noting property~(\ref{TheFormula}) and proceeding with
subsection \ref{KeyIdeas}, skipping the following subsection. \\

\subsection{Prefix Kolmogorov Complexity } \label{PrefixComplexity}

In this subsection we review the definition and some important
properties of the prefix complexity. Let
$\set{X}=\{\Lambda,0,1,00,01,10,11,000,\dots\}$ be the set of finite
binary strings where $\Lambda$ is the string of length 0.  Any subset
of $\set{X}$ is called a code. Any string in a code has a well defined
length and the set of string lengths is an important characteristic of
the code.  An instantaneous code is a set of strings $\set{Y}\subset
\set{X}$ with the property that no string in $\set{Y}$ is a prefix of
another.  A prefix computer is a partial recursive function\footnote{A
  partial function which is computed by a Turing machine.}  $C:
\set{Y}\times \set{X}\to \set{X}$.  For each $p\in \set{Y}$ (program
string) and for each $d\in \set{X}$ (data string) the output of the
computation is either undefined or given by $C(p,d)\in \set{X}$.
Following the usual motivation \cite{LiVitanyi}, we restrict our
attention to prefix computers.  This is a very weak restriction in the
sense that every uniquely decodable code can be replaced by an
instantaneous code without changing the set of string lengths
\cite{LiVitanyi}. Consider a mathematical object that has a binary
string $\alpha$ as its complete description. The idea is to choose
some reference computer $C$, find the shortest program that makes $C$
compute $\alpha$ given data $d$, and use the length $K_C(\alpha|d)$ of
the program (in bits) as the measure of the object's complexity.
Formally, the complexity of $\alpha$ given data $d$ relative to
computer $C$ is
\begin{equation}
K_C(\alpha|d)\equiv
\min_{p}\{ |p|\; {\big{|}}\;C(p,d)=\alpha\}\,,
\end{equation}
where $|p|$ denotes the length of the program $p$ (in bits).

Since this complexity measure depends strongly on the reference
computer, it is important to find an optimal computer $U$ for which
$K_U(\alpha|d)\leq K_C(\alpha|d)+\kappa_C$ for any
prefix computer $C$
and for all $\alpha$ and $d$, where $\kappa_C$ is a constant depending
on $C$ (and $U$) but not on $\alpha$ or $d$.  It turns out that the
set of prefix computers contains such a $U$ and, moreover, it can be
constructed so that any prefix computer can be simulated by $U$: for
further details consult~\cite{LiVitanyi}.  Such a $U$ is called a
universal prefix computer and its choice is not unique.  Using some
particular universal prefix computer $U$ as a reference, we define the
conditional Kolmogorov complexity of $\alpha$ given $\beta$ as
$K_U(\alpha|\beta)$.

The above definitions are generalized for the case of many strings
as follows. We choose and fix a particular recursive bijection $B:
\set{X}\times \set{X}\to \set{X}$.  Let $\{\alpha^i\}_{i=1}^{n}$ be a
set of $n$ strings $\alpha^i\in \set{X}$.  For $2\leq k\leq n\;$ we
define ${\langle\alpha^1,\alpha^2,\dots,\alpha^k\rangle}\equiv
B({\langle\alpha^1,\dots,\alpha^{k-1}\rangle},\alpha^k)$, and
${\langle\alpha^1\rangle}\equiv\alpha^1$.  We can now define
$K_U(\alpha^1,\dots,\alpha^n| \beta^1,\dots,\beta^k) \equiv
K_U({\langle \alpha^1,\dots,\alpha^n\rangle }|{\langle\beta^1,
  \dots,\beta^k\rangle})$.

For any two universal prefix computers $U_1$ and $U_2$ we have, by
definition, $|K_{U_1}(\alpha|\beta) -K_{U_2}(\alpha|\beta)| \leq
\kappa(U_1,U_2)$ where $\kappa(U_1,U_2)$ is a constant that depends
only on $U_1$ and $U_2$ and not on $\alpha$ or $\beta$. In many
standard applications of Kolmogorov complexity the set of reference
computers is considered to be finite and the attention is focused on
complex objects such as random or nearly random long strings. In such
cases, Kolmogorov complexity becomes an asymptotically absolute
measure of the complexity of individual strings: the constant
$\kappa(U_1,U_2)$ can  be neglected in comparison to the value of
the complexity.  For this reason, many fundamental properties of
Kolmogorov complexity are established up to an error term which can be
neglected compared to the complexity of the considered strings. For
instance, the standard analysis of the prefix Kolmogorov
complexity~(\cite{LiVitanyi}, Section 3.9.2) gives
\begin{equation}                        \label{ErrorTerm}
K_U(\alpha,\gamma|\beta)=K_U(\alpha|\gamma,\beta)
                                                       +K_U(\gamma|\beta)+\Delta\,,
\end{equation}
where $\Delta$ is the error term which grows logarithmically with the
complexity of considered strings. In our case such accuracy is
unacceptable as we want to use $K_U$ to analyze simple dynamical laws
for which the complexity is small and terms like $\Delta$ cannot be
neglected. Fortunately, in the case of simple strings (see Definition
1 in Ref.~\cite{SoklakovTCS}) this problem can be solved by a natural
restriction of reference computers \cite{SoklakovTCS}.  Roughly
speaking, this restriction entails the requirement that switching to a
more complex reference computer should always be accompanied by an
equivalent reduction of program lengths, i.e. more complex computers
are required to be more ``powerful''. Denoting a set of computers
which satisfies this requirement by $\{W_s\}$ we then construct a
computer $W$ which is universal for this set by setting $W(p,\langle
s,d\rangle)=W_s(p,d)$ and use any such $W$ as a reference. By a
slight abuse of notation, for any simple pair of strings 
$(\alpha,\gamma)$, we have by
Theorem~1 in Ref.~\cite{SoklakovTCS}:
 \begin{equation}                       \label{Kw}
K_W(\alpha,\gamma|\beta)=
     K_W(\alpha|\gamma,\beta)
   +K_W(\gamma|\beta)+{\mbox{\rm const}}\,,
\end{equation}
where the constant depends only on the reference machine $W$ (not on
$\alpha$, $\beta$ or~$\gamma$).

It is important to keep in mind that Kolmogorov complexity becomes a
particular function only if the reference computer is given. A set of
reference computers defines a set of complexity functions which have
some properties in common, e.g. Eq.~(\ref{Kw}), but nevertheless,
individual complexity functions can look very different from one
another. In order to verify whether any particular function
$G:\set{X}\to\set{N}$ is a Kolmogorov measure of complexity it is
necessary and sufficient to find a reference computer $W$ such that
$K_W=G$. The rest of this subsection deals with the properties that
are common to all complexity functions defined by the set of reference
computers $\{W\}$.
 
For any particular reference computer $W$, we can simplify the
notation $K\equiv K_W$ and use (\ref{Kw}) to show that
\begin{eqnarray} \label{2}
K(\alpha^1,\dots,\alpha^N|d)&=&K(\alpha^N|d)
        +K(\alpha^1,\dots,\alpha^{N-1}|\alpha^N,d)
                       + {\rm const}   \cr
                            &=& \cdots \cr
                            &=&K(\alpha^N|d)
+\sum_{n=1}^{N-1}K(\alpha^{N-n}|\alpha^{N-n+1},\dots,
                            \alpha^N,d)
                         +{\rm const} \cr
                            &=&K(\alpha^N|d)
+\sum_{n=1}^{N-1}K(\alpha^{n}|\alpha^{n+1},\dots,
                          \alpha^N,d)
                         +{\rm const}\;.
\end{eqnarray}
Defining the conditional mutual information of objects $\alpha$ and
$\gamma^1,\gamma^2,\dots,\gamma^N$ as
\begin{equation}
I(\alpha:\gamma^1,\dots,\gamma^N|d)\equiv K(\alpha|d)
-K(\alpha|\gamma^1,\dots,\gamma^N ,d)
\end{equation}
(consult Ref.~\cite{LiVitanyi}) we have
\begin{equation} \label{TheFormula}
K(\alpha^1,\dots,\alpha^N|d)=
              \sum_{n=1}^{N}K(\alpha^n|d)
              -\sum_{n=1}^{N-1}
              I(\alpha^n:\alpha^{n+1},\dots,\alpha^N|d)
              + {\rm const}.
\end{equation}
This equation will soon become important for the complexity analysis
of dynamical laws. \\

\subsection{Complexity of dynamical laws } \label{DynComplexity}

Recall that a dynamical law was defined earlier as a function on the
state space of the system. If we had a definition of the complexity of
a function we would therefore be able to
quantify the complexity of a dynamical law.  For any
function $\Jf: \set{X}\to \set{X}_1\times\cdots\times\set{X}_J$
$(\set{X}_j=\set{X})$,  the complexity of  
$\Jf$ at $x_0\in\set{X}$ is defined as
\begin{equation} \label{KFunc}
K_{x_0}[\Jf]\equiv\frac{1}{J} \sum_{k=0}^{J-1} K(x_{k+1}|x_k)\,,
\end{equation}
where the ordered sequence $\{x_1,x_2,\dots,x_J\}=\Jf(x_0)$.
It is helpful to illustrate this definition for the case of Markovian
laws as defined in section~\ref{sec:SP}.  
If $\Jf$ is Markovian, then by definition
there exists a function $g: \set{X} \to \set {X}$ such that $x_{k+1}=g(x_k)$
and we have
\begin{equation} \label{KFunc2}
K_{x_0}[\Jf]=\frac{1}{J} \sum_{k=0}^{J-1} K_{x_k}[g]\,.
\end{equation}
The complexity of $\Jf$ is therefore equal to the complexity of $g$
at a typical point in the trajectory $\{x_0,\dots,x_J\}$. In other
words, the complexity of $\Jf$ at $x_0$ quantifies the amount
of information needed to compute a typical step of the trajectory
generated by $\Jf$ from the initial condition $x_0$. \\

\subsection{Key ideas} \label{KeyIdeas}
Before we introduce our derivation of Newtonian mechanics, it is
relevant to recall the definition of a Newtonian mechanical system and
highlight the conceptual difficulties of the standard approach.
A Newtonian mechanical system consists of particles
whose dimensions can be neglected in describing their motion. The
position of a particle in space is defined by its Cartesian
coordinates $\br=(x,y,z)$. The derivative
$\dot{\br}=(\dot{x},\dot{y},\dot{z})\equiv (dx/dt,dy/dt,dz/dt)$ of the
coordinates with respect to time $t$ is called the Cartesian
velocity of the particle. The physical state of the system is
completely determined if the coordinates and the velocities are
determined for every particle in the system. For every mechanical
system one can write a function of its state that together with an
appropriate dynamical principle defines the system evolution. This
function is called the Lagrangian of the system and is usually
postulated, except for a few special cases where it can be derived.
For example, one can show that the Lagrangian of a single free
particle is proportional to its squared velocity. This fact is a
direct consequence of the Galilean principle of relativity and the
classical definitions of homogeneous isotropic space and homogeneous
time (Ref.~\cite{LandauM}, \S 3,4).  Unfortunately, this is about all
one can explain using the standard approach.
Certainly, we have no satisfactory explanation of why
the Lagrangian of a mechanical system has the accepted form $L=T-V$
and why we minimize a functional of ``action'' which is an integral of
$L$ along a short segment of a path.  
Indeed, the standard derivation of the equations of motion from
the principle of stationary action uses Newton's second law as an
established fact \cite{Goldstein}. An analogous situation is found in
the standard derivation of the equations of wave mechanics in the
Lagrangian formulation \cite{Ryder}. It may seem that the Lagrangian
formulation of the laws of dynamics is merely one way of writing them
down. Nevertheless, the Lagrangian formulation plays an important role
in {\it understanding} the physical world in many areas due to its
truly remarkable ability of unifying various types of interactions.
Quoting R.~P.~Feynman \cite{Feynman} ``We regard the action to be the
more fundamental quantity.  From it we can immediately read off the
rules for the propagators, the coupling, and the equations of motion.
{\it But we still do not know the reason}\footnote{Italics introduced
  by this author and not in the original text.} for the rules for the
diagrams, or why we can get the propagators out of $\ttS$ [the
action]''. Whatever the physical interaction, if it is well defined
and understood, it is often enough to add one extra term in the
Lagrangian to describe it.

Adding interaction terms to free Lagrangians is a rather specific
way of introducing interactions. In conservative
nonrelativistic mechanics, for example, interactions are typically considered
as functions of the {\it relative} positions 
of the interacting subsystems~(Ref.~\cite{LandauM}, \S 5).
Lagrangians of 
free subsystems are functions of a different type: they
can only depend on the {\it absolute} states describing each
subsystem individually. The total Lagrangian, including
interaction, is constructed as a difference between free
Lagrangians and interaction terms. Not every function
of the combined system state can be represented in this
way. 

We suggest that property (\ref{TheFormula}) of
complexity measures may provide an explanation for the general structure 
of the Lagrangian for a composite physical system.
Considering, for instance, a pair of strings $\alpha^1$ and
$\alpha^2$, we have from (\ref{TheFormula}) that the complexity of
them both given any data $d$ is given by
\begin{equation}                        \label{BipartiteK}
K(\alpha^1,\alpha^2|d)=K(\alpha^1|d)+K(\alpha^2|d)
-I(\alpha^1:\alpha^2|d)+{\rm const}\;, 
\end{equation}
where the first two terms represent the complexities of $\alpha^1$ and
$\alpha^2$ considered independently from one another.  The third term
$I(\alpha^1:\alpha^2|d)$ quantifies the strength of correlation
between the two strings which can be viewed as an amount of
information in one string about the other, given initial knowledge
$d$.  A typical action of a composite physical system has the same
structure. For the action $\ttS^{1,2}$ of a bipartite system we would
normally write
\begin{equation}                        \label{BipartiteS}
\ttS^{1,2}=\ttS^1+\ttS^2-\ttS^{\rm int} +{\rm const}\;,
\end{equation}
where $\ttS^1$ and $\ttS^2$ are the actions for individual subsystems,
$\ttS^{\rm int}$ is the interaction term, and the constant can be
arbitrary.  The similarity between (\ref{BipartiteK}) and
(\ref{BipartiteS}) becomes even more apparent if we think of
correlation between two strings $\alpha^1$ and $\alpha^2$ as a
manifestation of interaction.  To be more precise, we can use the
strings $\alpha^1$ and $\alpha^2$ to describe dynamical laws $g^1$ and
$g^2$ of individual subsystems by setting $\alpha^k=g^k(d)$.
Using (\ref{BipartiteK}) and the definition of complexity
of a dynamical law (section~\ref{DynComplexity}), we have
\begin{equation}                        \label{K_f1_f2}
K_d[g^1,g^2]=K_d[g^1]+K_d[g^2]
-I_d[g^1:g^2]+{\rm const}\;, 
\end{equation}
where $I_d[g^1:g^2]\equiv I(\alpha^1:\alpha^2|d)$ quantifies the
strength of correlation between the two dynamical laws governing the
interacting subsystems. We draw special attention to the fact that the
property given by Eq.~(\ref{BipartiteK}) is rather typical for
information-theoretic measures of complexity and can be easily
understood using Venn diagrams. We can therefore anticipate that the
structure of action for a composite physical system (\ref{BipartiteS})
can be understood as a consequence of a more general property, namely
the structure of complexity of dynamical laws governing the behaviour
of the system. 

In this article, we develop the proposed approach considering a
particular measure of complexity, namely Kolmogorov complexity. Our
choice of this measure is based on the fact that Kolmogorov complexity
was specifically designed for quantification of the
complexity of individual objects, as opposed to alternative
probabilistic approaches. The price we pay is a rather difficult or,
more likely, unusual mathematical formalism. Kolmogorov complexity is
defined with respect to a ``reference computer''; it is an essentially
discrete quantity; and there is no algorithm which can compute
this quantity in the most general case.  It is hard to imagine a more
difficult quantity in the realm of physics where we are used to
integrable, differentiable or even smooth functions.

We have already mentioned that the Galilean relativity principle plays
a key role in the derivation of the Lagrangian of a single nonrelativistic
particle but alone it is not enough to explain all the postulates of the
Hamilton's principle of least action. Using a simple example
of a conservative nonrelativistic system we show that the Galilean
relativity principle can be combined with the {\it SP} to answer
the questions posed at the beginning of this article.
The structure of the complexity of the dynamical laws
given by Eq.~(\ref{K_f1_f2}) explains the general structure
of the Lagrangian of a composite system; the integral in the
definition of the action corresponds to the sum in Eq.~(\ref{KFunc2});
and the minimization procedure corresponds to finding the simplest
dynamical law (consistent with the Galilean relativity).
These arguments only deal with the structure of the Hamilton's
principle and can be applied beyond nonrelativistic mechanics.
In the relativistic case, for instance, the same arguments
apply, the only difference being that the Galilean relativity principle
is replaced with the relativistic principles of
Einstein. 

 It can already be seen from this introduction that the
invariance of action associated with a particular relativity principle
must be satisfied by the measure of complexity of the dynamical laws.
This is the point where the technical difficulties associated with the
use of Kolmogorov complexity appear.
In principle, by choosing an appropriate reference computer,
it is possible to set the Kolmogorov complexity to any function
at any {\it finite} set of strings. The problem is that it can be difficult
to construct a ``natural'' example of such a computer.
The good news is that the relativity principles have nothing
to do with the properties of action like the structure~(\ref{BipartiteS}).
This means that in order to check the
consistency of the {\it SP} with a particular relativity principle
it is sufficient to consider only the free particle case.

In our derivation of Newton's second law
we construct a rather artificial example of a reference computer which
satisfies all the constraints imposed by the Galilean relativity principle
on the complexity of physical laws. It is interesting,
however, that the complications encountered in the case of
Newtonian mechanics disappear in the relativistic case, even though the
arguments are absolutely analogous. Mathematically, this is due to the
fact that the square of the four-velocity of a relativistic particle
is always equal to one, whereas the squared velocity of a
nonrelativistic particle, which appears in the Lagrangian, depends on
the reference frame.  It is tempting to assume that the special
role of time in Newtonian mechanics is to blame for the complications.
At the moment, however, there is no evidence that the reference
computer in the nonrelativistic case cannot be constructed in a more
elegant way. 

In conclusion of this section, it is important to emphasize
that, in this article, we use only a small fraction of
Kolmogorov complexity calculus. Kolmogorov complexity 
is rich in properties which can be useful in fundamental physics. 
As a simple example, consider the Kraft inequality which
demands that the sums of the type $\sum_f2^{-K_x[f]}$
are convergent. The proposed analogy between Kolmogorov
complexity and action suggests that the Kraft inequality
may be useful in the context of the path integral approach
(see section~\ref{Discussion} for some details). \\

\section{Main derivations }  \label{main}

Given a physical system, consider the set $\set{S}$ of all possible
states of the system. For any initial state $\xi_0\in\set{S}$ and for
any dynamical law $f$ the entire system evolution is given by the
trajectory $f(\xi_0)=\{\xi_1,\xi_2,\dots\}$ in the ``state space''
$\set{S}$ of the system. This definition is general enough in that 
a continuous trajectory can be defined as a sequence
of points $\xi_s$, where $s$ is a continuous parameter.
In the case of a composite system, one can
also introduce state spaces and dynamical laws for every subsystem.
In general, these dynamical laws are not independent but correlated
due to physical interaction between the subsystems.  One way to study
that correlation in detail is the complexity analysis based on the
earlier defined notion of Kolmogorov complexity of a function.  In
this case a coarse-graining of the state space is often necessary as
we need finite binary strings $\bar{\xi}_k$ to address the points
$\xi_k$ in $\set{S}$ which is often a continuum.  It is convenient to
identify binary strings with the coarse-grained numerical values they
represent.  One can also assume, without loss of generality, that the
coarse-graining of the state space can be performed as fine as
necessary at the cost of increasing the length of the binary strings
used.

Let a sequence of finite binary strings
$\bar{\xi}_0,\bar{\xi}_1,\dots,\bar{\xi}_J$ represent some $J+1$
points of a coarse-grained trajectory in the state space.  A function
$\Jf:\set{S}\to \set{S}_1\times\cdots \times\set{S}_J$ $(\set{S}_j=\set{S})$
 of the collective
variable $\bar{\xi}$ defines a coarse-grained dynamical law if
\begin{equation}
\Jf(\bar{\xi}_0)
=\{\bar{\xi}_1,\bar{\xi}_2,\dots,\bar{\xi}_J\}.
\end{equation}
In the standard approach we consider a set of continuous
differentiable trajectories with fixed initial and final conditions.
Such trajectories can obviously be approximated to any degree
of accuracy by the $\Jf$-type of functions. To get a better
approximation all we need to do is to increase the length
of binary strings, keep $\bar{\xi}_0$
and $\bar{\xi}_J$ fixed while adding more points
in between. Formally we write
\begin{equation}                        \label{continuous}
f=
\lim_{J\to\infty}\lim_{|\bar{\xi}|\to\infty} \Jf\,,
\end{equation}
where the limit indicates that the continuous trajectory $f(\xi_0)$
has an infinite number of points $(J\to\infty)$ each specified  to an
infinite accuracy ($|\bar{\xi}|\equiv\sum_k |\bar{\xi}_k|\to\infty$).
In any situation when we are interested in continuous laws,
we would normally start from the continuous trajectory
$f(\xi_0)$ so there is no ambiguity in the definition of the limit. 

According to the definition of complexity of a function (subsection\ 
\ref{PrefixComplexity}), the complexity of a coarse-grained dynamical law
$\Jf$ is given by
\begin{equation}                              \label{LimEasyLearnable}
K_{\bar{\xi}_0}[\,\Jf\,]=
\frac{1}{J}\sum_{k=0}^{J-1}
K(\bar{\xi}_{k+1}|\bar{\xi}_{k})\,.
\end{equation}
The {\it SP} can now be formulated as a variational problem
of minimizing $K_{\bar{\xi}_0}[\Jf]$ over all dynamical laws
$\{\Jf\}$ that are consistent with all other physical axioms.
It is important to acknowledge that the {\it SP} is just an inference
tool and additional axioms are needed for finding physical dynamics.
It is for this reason that we cannot narrow the set of dynamical laws
down to the trivial law
$(\br_{k+1},\dot{\br}_{k+1})=(\br_{k},\dot{\br}_{k})$, which is
intuitively the simplest. In Newtonian mechanics, for instance, the
minimal set of axioms includes the {\it SP} {\it and} the Galilean
relativity principle. The trivial law would violate the Galilean
relativity principle as a particle cannot be at rest in more than one
reference frame. \\

\subsection{Newtonian Mechanics } \label{NewtonianMechanics}

Consider the state space of a Newtonian mechanical system where time,
positions and velocities of the particles are combined into a
collective variable{}\footnote{Here we omit the subscript enumerating
  the particles for the sake of convenience of notation.}  $\xi\equiv
(\br, \dot{\br}, t)$ to represent every possible state of the system
\cite{Lanczos}.  Each point $\bar{\xi}_k$ of a coarse-grained
trajectory approximates a point $\xi_k\equiv (\br_k, \dot{\br}_k,
t_k)$ in the state space. We can therefore define $\bar{\xi}_k\equiv
{\langle\bar{\br}_k,\oo{\br}_k,\bar{t}_k\rangle}$ where $\bar{\br}_k$,
$\oo{\br}_k$ and $\bar{t}_k$ are finite binary strings approximating
the values of $\br_k$, $\dot{\br}_k$ and $t_k$ in the real-parameter
state space.

We shall follow the standard approach and choose the time $t\equiv
\bar{t}_k$ instead of an abstract parameter $k$ to define the order of
events $\{\bar{\xi}_k\}$ in the trajectory $\Jf(\bar{\xi}_0)$.  This
reflects the absolute nature of time in Newtonian mechanics and is not
a necessary requirement of our approach.  We put all elements of the
set $\{t\}$ in order such that for any two consecutive times $t$ and
$t+\Delta t$ we have $\Delta t>0$.  For any function of time $\phi_t$
we define $\Delta\phi_t\equiv\phi_{t}-\phi_{t-\Delta t}$ and the {\it
  discrete time derivative} is defined as
\begin{equation}                        \label{DiscreteTimeDerivative}
\oo{\br}_t\equiv \frac{\Delta\bar{\br}_t }{\Delta t}
 =\frac{ \bar{\br}_{t}-\bar{\br}_{t-\Delta t}  }{  \Delta t   }\;.
\end{equation} 
We see that if $|\bar{\xi}|\equiv\sum_k|\bar{\xi}_k|$ is taken to
infinity one can chose the lengths of $\{\bar{\br}_k\}$ and
$\{\bar{t}_k\}$ such that $\oo{\br}_t$ approaches $\dot{\br}_t$. This
fact is used every time we approximate derivatives by ratios of finite
differences on a computer. Formally we write
\begin{equation}                        \label{ContinuumLimit}
\dot{\br}_t=
\lim_{\Delta t\to 0}\lim_{|\bar{\xi}|\to\infty}\oo{\br}_t\;,
\end{equation}
where, as in the case of Eq.~(\ref{continuous}), the double limit means
that, in practice, sufficiently long binary strings ($|\bar{\xi}|\to
\infty$) should be used for any finite $\Delta t$ (i.e.,\ longer
strings are needed for better precision). Equation
(\ref{ContinuumLimit}) suggests that the above definition of
$\oo{\br}_t$ can be used for construction of $\bar{\xi}_t\equiv
{\langle\bar{\br}_t,\oo{\br}_t, t\rangle}$ which is by definition a
coarse-grained approximation of $\xi_t=
{\langle\br_t,\dot{\br}_t,t\rangle}$.  The second discrete time
derivative is, by analogy, defined as
\begin{equation}                        \label{IIDiscreteTimeDerivative}
\ooo{\br}_t\equiv\frac{\Delta\oo{\br}_t}{\Delta t}
     =\frac{ \oo{\br}_{t}-\oo{\br}_{t-\Delta t}   }{  \Delta t   }\;.
\end{equation}

In the new parameterization Eq.~(\ref{LimEasyLearnable}) becomes
\begin{equation}                        \label{NewParametrization}
K_{\bar{\xi}_0}[\Jf]=
    \frac{1}{J}\sum_{t=\bar{t}_0}^{\bar{t}_{J-1}}
   K(\bar{\br}_{t+\Delta t},\oo{\br}_{t+\Delta t}, t+\Delta t |
                 \bar{\br}_t,\oo{\br}_t, t)
                               \;,
\end{equation}
where the sum over $t$ goes through the set
$\{\bar{t}_k\}_{k=0}^{J-1}$; we choose for simplicity
$t\equiv\bar{t}_k=k\Delta t$, where $\Delta t$ is now a constant. Like
any other re-parameterization, this relation can be absorbed into the
definition of the reference computer in the form of a subroutine which
is always executed to calculate $t+\Delta t$ given $t$. Equation
(\ref{NewParametrization}) becomes
\begin{equation}                        \label{AbsorbTPlusDeltaT}
K_{\bar{\xi}_0}[\Jf]  =
    \frac{1}{\bar{t}_J-\bar{t}_0}\sum_{t=\bar{t}_0}^{\bar{t}_{J-1}} 
   K(\bar{\br}_{t+\Delta t},\oo{\br}_{t+\Delta t} |
                 \bar{\br}_t,\oo{\br}_t, t)\, \Delta t
                               \;.
\end{equation}
It is also convenient to absorb the definitions of discrete time
derivatives, Eqs.~(\ref{DiscreteTimeDerivative}) and
(\ref{IIDiscreteTimeDerivative}), into the definition of the reference
computer. In this case $\oo{\br}_{t+\Delta t}$ and
$\bar{\br}_{t+\Delta t}$ can be set for automatic evaluation from
$\ooo{\br}_{t+\Delta t}$ given $\oo{\br}_t$ and $\bar{\br}_t$. We
therefore have
\begin{equation}
K_{\bar{\xi}_0}[\Jf]=
\frac{1}{\bar{t}_J-\bar{t}_0}\sum_{t=\bar{t}_0}^{\bar{t}_{J-1}}
   K(\ooo{\br}_{t+\Delta t} |
                 \bar{\br}_t,\oo{\br}_t, t)\, \Delta t
                               \;.
\end{equation}
Let us fix the time $\tau\equiv t_J-t_0$ in which we investigate the
system evolution. We have
\begin{equation}                        \label{IntroduceTau}
K_{\bar{\xi}_0}[\Jf]=\frac{1}{\tau}
                  \sum_{t=0}^{ \tau-\Delta t}
   K(\ooo{\br}_{t+\Delta t} |
                 \bar{\br}_t,\oo{\br}_t, t)\, \Delta t
                               \;,
\end{equation}
where the sum over $t$ goes from $0$ to $\tau-\Delta t$ in steps of
$\Delta t$.  From this equation we can already see that the complexity
of dynamical laws is determined by the complexity of the acceleration
$\ooo{\br}_{t+\Delta t}$ given the system state $\bar{\xi}_t=
{\langle\bar{\br}_t,\oo{\br}_t, t\rangle}$ in the immediately
preceding past.

Consider the simple example of a single free particle which at the
instants $0$ and $\tau$ is in the states $\bar{\xi}_0$ and
$\bar{\xi}_\tau$ respectively.  The physical dynamics which are to be
found by minimizing $ K_{\bar{\xi}_0}[\Jf]$ over all $\Jf$
with fixed $\bar{\xi}_0$ and $\bar{\xi}_\tau$ should satisfy
the Galilean relativity principle.
This restricts the set of dynamics $\{\Jf\}$ to
those for which
\begin{equation}                        \label{OneParticle}
 K(\ooo{\br}_{t+\Delta t}|\bar{\br}_t,\oo{\br}_t, t )
=\frac{m\oo{\br}{}^2_t}{2} +\frac{\Delta Q(\bar{\xi}_t) }{ \Delta t}\;,
\end{equation}
where $m$ is a positive coefficient and $Q(\bar{\xi}_t)$ is an
arbitrary function of the state of the system. Before we proceed with
the proof of this, it is relevant to recall the Galilean relativity
principle.

The Galilean relativity principle is based on the notion of {\it
  inertial reference frames} in which, by definition, the laws of
mechanics take their simplest form (Ref.~\cite{LandauM}, \S 3).
Mathematically, inertial reference frames are defined through a number
of properties which are known commonly as Newton's First law. These
properties imply homogeneity and isotropy of space and homogeneity of
time. Moreover, in the inertial reference frame where a free body is
at rest at some instant it remains always at rest.  And finally, the
coordinates $\br$ and $\br'$ of a given point in two different
inertial frames are related by the Galilean transform
\begin{equation}
\br=\br'+\bv t\;,
\end{equation}
where it is understood that time is the same in the two frames
($t=t'$) and that the second frame moves relative to the first one
with velocity $\bv$. It follows directly from these definitions that
velocity of a free particle is constant in any inertial frame, i.e.\ 
$\ooo{\br}_{t+\Delta t}=0$.

The proof of equation~(\ref{OneParticle}) is constructed as follows.
Suppose we want to interpolate $K(\ooo{\br}_{t+\Delta t} |
\bar{\br}_t,\oo{\br}_t, t )$ by some well behaved function $L_1$ which
is defined on real numbers, but coincides with $K(\ooo{\br}_{t+\Delta
  t} |\bar{\br}_t,\oo{\br}_t, t )$ on the coarse-grained trajectory
where $K$ is defined.  Assuming that such an $L_1$ exists, we
determine its properties and prove that on the coarse-grained state
space it would behave as suggested by Eq.~(\ref{OneParticle}). We then
show that there exist infinitely many reference computers for which
$K$ is consistent with Eq.~(\ref{OneParticle}). In such cases the
interpolation $L_1$ of $K$ can be found as assumed,
because it is
uniquely (up to the total time derivative) defined by
Eq.~(\ref{OneParticle}).  In summary, we demonstrate that by choosing
an appropriate reference computer $W$, the complexity measure $K$ can
be made to satisfy all the requirements imposed by the Galilean
relativity principle in the framework of the {\it SP}.
In this argument we use the
standard formulation of the Galilean relativity principle on the
continuum. Alternatively, we could reformulate the Galilean relativity
principle in a discrete form and try to apply it directly to $K$
without introducing $L_1$. This will be considered in future research.
However for now, we shall keep the standard formulation of the
Galilean relativity principle on the continuum, and demonstrate the
relation between the continuous and the discrete formulations of
Newtonian mechanics. Later we will see that in the relativistic case
all the arguments can be performed without introducing $L_1$.

For our case of a single free particle we have $\ooo{\br}_{t+\Delta
  t}=0$, and therefore $L_1$ is a function of only two vector and one
scalar arguments (coordinates $\br_t$, velocity $\dot{\br}_t$ and
time~$t$).  Substituting such an $L_1$ instead of $K$ into
Eq.~(\ref{IntroduceTau}) we have
\begin{equation}                        \label{IntroduceL1}
K_{\bar{\xi}_0}[\Jf]=
                  \frac{1}{\tau}
                  \sum_{t=0}^{ \tau-\Delta t}
   L_1(\bar{\br}_t,\oo{\br}_t, t) \Delta t
\end{equation}
We require integrability of $L_1$ on $\set{R}$, in which case
the complexity of continuous dynamics~(\ref{continuous}) 
can be quantified by
\begin{eqnarray}
S[f]&\equiv&  \frac{1}{\tau}\lim_{\Delta t\to 0}\lim_{|\bar{\xi}|\to\infty}
                           \sum_{t=0}^{ \tau-\Delta t}
                           L_1(\bar{\br}_t,\oo{\br}_t, t) \Delta t\cr
&=&\frac{1}{\tau}\int_0^ \tau L_1(\br_t,\dot{\br}_t, t)\, d t
                               \;.
\end{eqnarray}  
Because of the formal connection of $S[f]$ to the physical laws
through the minimization procedure we shall call $L_1$ the Lagrangian
and $S[f]$ the corresponding action for a single free particle.
Following the argument by Landau and Lifshitz (\cite{LandauM}, \S
3,4), we note that the homogeneity and isotropy of space and
homogeneity of time in an inertial reference frame imply that the
Lagrangian $L_1$ can only depend on the absolute value of velocity
\begin{equation}
L_1(\br,\dot{\br}, t)=L_1(\dot{\br}^2)+\frac{d}{dt}\tilde{Q}(\xi_t)\;.
\end{equation}  
Here the total time derivative is introduced to emphasize that with
fixed initial and final conditions the variational problem of
minimizing $S[f]$ over all $f$ is not affected by addition of any
$d\tilde{Q}(\xi_t)/dt$ to the Lagrangian $L_1(\br,\dot{\br}, t)$.  The
Galilean relativity principle requires that in the reference frame
which moves with infinitesimal velocity $\bepsilon$ relative to the
original inertial reference frame the Lagrangian $L_1(\dot{\br}'{}^2)
=L_1(\dot{\br}^2+2\dot{\br}\cdot\bepsilon+\bepsilon^2)$ can differ
from $L_1(\dot{\br}^2)$ only by the total time derivative of some
function of the particle state. This implies that $\partial
L_1/\partial\dot{\br}^2$ does not depend on the velocity, because the
second term in the expansion
\begin{equation}
L_1(\dot{\br}'{}^2)=L_1(\dot{\br}^2)
  +\frac{\partial L_1}{\partial \dot{\br}^2}2\dot{\br}\cdot\bepsilon
                   +\dots
\end{equation} 
is a total time derivative only if it is a linear function of
$\dot{\br}$.  Writing $\partial L_1/\partial\dot{\br}^2=m/2$ and
neglecting the total time derivative we have
\begin{equation}
  L_1(\br,\dot{\br}, t)=\frac{m\dot{\br}^2}{2}.
\end{equation}
Because the variational problem of minimizing $K_{\bar{\xi}_0}[\Jf]$
with fixed
initial and final conditions is not affected by addition of $\Delta
Q(\bar{\xi}_t)/\Delta t$ to the complexity $K(\ooo{\br}_{t+\Delta
  t}|\bar{\br}_t,\oo{\br}_t, t ) = L_1(\bar{\br}_{t+\Delta
  t},\oo{\br}_t, t)$, the above equation implies
Eq.~(\ref{OneParticle}) as required. The standard derivation of the
free particle Lagrangian $L_1$ used here gives equivalent results in
any coordinate system: $L_1$ is always the kinetic energy.  Likewise, our
requirements on complexity are physically equivalent for different
parameterizations $\{\bar{\xi}\}$ of the coarse-grained state space.  The
Galilean relativity principle demands~(\ref{OneParticle}), so that the
choice of parameterization of the state space does not matter.

It now remains to show that the constraint on the form of the
complexity imposed by the Galilean relativity principle can be
satisfied by a considerable number of reference computers. Certainly,
not every reference computer $W$ would satisfy this. It is however not
surprising because the definition of $W$ is far too general.  To prove
that there is a $W$ for which (\ref{OneParticle}) is true, imagine
that we have found the physical dynamical law.  For this particular
law there is a corresponding computer which by default (if given zero
length program) calculates the physical value of $\ooo{\br}_{t+\Delta
  t}$ from any ${\langle\bar{\br}_t,\oo{\br}_t, t\rangle}$. For such a
computer, the minimal value of complexity $K(\ooo{\br}_{t+\Delta
  t}|\bar{\br}_t,\oo{\br}_t, t)=0$ is obtained for the physical law.
This is achieved in the fixed reference frame ${\cal F}_0$ where the
physical law was found. In order to perform computations in any given
reference frame, we modify the computer to wait for a string of code
which is appended to the main program.
This appended code accommodates
the definition of the given reference frame relative to the fixed
${\cal F}_0$, and reformulates the results of the main program in the
given frame.  We choose the fixed frame ${\cal F}_0$ as the rest frame
of our free particle, and require that the appended code has the fixed
length of $m\oo{\br}{}_t^2/2+ c$, where $c$ is a constant.
The computer would read
$\oo{\br}_t$ from the given data and perform calculations only if the
appended code has the required length. Because $c$ can be rather big
the shortest program describing a dynamical law $\Jf$ would use
the unnecessary space in the appended code to encode as much
information about $\Jf$ as possible. In fact, for simple $\Jf$ the length
of the shortest program will coincide with the length of the appended
code. We therefore constructed a computer which satisfies
requirement (\ref{OneParticle}) for simple dynamical laws
and, moreover, because we can choose
$c$ in a countably infinite number of ways, we know that
there are at least a countably infinite number of computers which
satisfy Eq.~(\ref{OneParticle}).  This construction, although
mathematically consistent, is rather artificial as the length of the
appended code explicitely depends on the system state.  This
unpleasant feature disappears in the relativistic case where the
appended code is not necessary. This completes the consideration of
the case of a single nonrelativistic particle. The same arguments can
be used to study the system of $N$ noninteracting particles because
they can be considered independently from one another.

To study the case of $N$ interacting particles we split the total
system state space $\set{S}=\{\bar{\xi}_t\}$ into subspaces
$\set{S}^{i}=\{\bar{\xi}^i_t\} $ that correspond to individual
particles so that
$\set{S}^1\times\set{S}^2\times\dots\times\set{S}^N=\set{S}$.  One
easy way to do this is to construct a state space for every particle
$\set{S}^i=\{\bar{\xi}^i_t\}$ and then use the bijection $B$ to form
the total state space as ${\set{S}=\{\bar{\xi}_t\ |\ \bar{\xi}_t
  ={\langle\bar{\xi}^1_t,\dots,\bar{\xi}^N_t\rangle}\}}$.  Using
(\ref{TheFormula}), equation (\ref{IntroduceTau}) for the entire
system can be rewritten as
\begin{eqnarray}
K_{\bar{\xi}_0}[\Jf]&=&\frac{1}{\tau}
   \sum_{t=0}^{\tau-\Delta t}
   K(\ooo{\br}{}^1_{t+\Delta t},\dots,\ooo{\br}{}^N_{t+\Delta t}|\bar{\xi}_t)
           \,\Delta t \cr
    &=& \frac{1}{\tau}
\sum_{t=0}^{\tau-\Delta t}
     \left(    \sum_{n=1}^N
                      K(\ooo{\br}{}^n_{t+\Delta t}|\bar{\xi}_t)
       -V_N+E\right)\Delta t\;.
\end{eqnarray}  
where $E$ is a constant of motion and, up to the additive total time
derivative,
\begin{equation}                        \label{Vn}
V_N\equiv\sum_{n=1}^{N-1}
                I(\ooo{\br}{}^n_{t+\Delta t}:\ooo{\br}{}^{n+1}_{t+\Delta t},
                                   \dots,\ooo{\br}{}^{N}_{t+\Delta t}|\bar{\xi}_t)      \;.
\end{equation}
We have already mentioned that any interaction between subsystems
manifests itself in the correlation of their dynamics. This
correlation is quantified by the mutual information and, for this
reason, we will call $V_N$ the interaction term. Function (\ref{Vn})
obviously contains the Newtonian case of binary interaction
$V_N=\sum_{j<k} V_{jk}$ where $V_{jk}$ stands for the interaction
between particles $j$ and $k$. In subsequent work we will present a
more detailed study of $V_N$. Here we suppose that interaction between
subsystems is known from experiment so that the interaction term is
given as a function on the state space of the system. For simplicity
we will consider the case when $V_N$ is a function only of the
coordinates
$\{\bar{\br}_t^n\}_{n=1}^N$. Moreover, it would be misleading
to consider a more general case, since velocity dependent forces
appear in mechanics as an attempt to include friction
or electro-magnetic interactions. Friction is essentially
an {\it effective} phenomenon: that is, there is no single fundamental
interaction which describes friction without further approximations.
 Electro-magnetic interactions are also
irrelevant as they are not Newtonian.
For these reasons, velocity independent potentials are used
as a standard requirement for fundamental derivations
in nonrelativistic mechanics (Ref.~\cite{LandauM}, \S 5).

The question now is to determine $K(\ooo{\br}{}^n_{t+\Delta
  t}|\bar{\xi}_t) $.  During a sufficiently short interval of time
$\Delta t$, the velocities of the particles can be treated as
constants, which is analogous to the case of zero interaction.  One can
therefore repeat the arguments as in the case of
Eq.~(\ref{OneParticle}), with the reservation that the coefficient $m$
can depend on time and can be different for every particle in the
system. Physically, this would correspond to the most general case
when particles' masses are changing with time like, for instance, in
jet motion. Mathematically, this possibility arises because the
arguments should be repeated for every short interval of time $\Delta
t$ independently. In doing so we should treat $\ooo{\br}{}^k_{t+\Delta
  t}$ as a constant because it relates asymptotically constant
velocities of the particles during different intervals $\Delta t$.
As explained above, for sufficiently small $\Delta t$, we have:
\begin{equation}                        \label{Sf}
K_{\bar{\xi}_0}[\Jf]=
 \frac{1}{\tau}
\sum_{t=0}^{\tau-\Delta t}
     \left(    \sum_{n=1}^N \frac{m_n(\oo{\br}{}_t^n)^2}{2}                      
       -V_N+E\right)\Delta t\;.
\end{equation}
Discrete formulation of Newton's second law appears as a necessary
condition for the variational problem of minimizing $K_{\bar{\xi}_0}[\Jf]$,
which for the case of one particle moving in a potential gives, as shown in the
Appendix ,
\begin{equation}                        \label{DiscreteNewton}
\frac{\Delta}{\Delta t} [m(\oo{\br}_t+\delta\oo{\br}_t/2 )]=
-\frac{\delta V(\bar{\br}_{t-\Delta t})}{\delta\bar{\br}_{t-\Delta t}}\;.
\end{equation}
The right hand side of this equation is a discrete variational
derivative as defined in the Appendix . We see that the acceleration
is determined by the force acting on the particle in the immediately
preceding past. We can therefore conclude that the force is the cause
of acceleration in the inertial reference frame. Taking the continuum
limit $|\bar{\xi}|\to\infty$ followed by $\Delta t\to 0$ we recover
the standard formulation of Newton's second law
\begin{equation}
\frac{d}{dt}(m\dot{\br})=-\frac{d V(\br)}{d\br}\;.
\end{equation}
Note that for the investigation of causality and related topics such
as the arrow of time, our discrete formulation of Newton's second law
is better suited than the standard formulation. We shall leave these
topics for further research and proceed with important special cases
where the difference between the discrete and the differential
formulations of dynamical laws can be neglected. In these cases the
differential form of the dynamical laws can be determined by applying
techniques of standard variational calculus for minimizing the
functional
\begin{equation}                        \label{As}
A_S\equiv a\int_{0}^{\tau} 
\left(\sum_{n=1}^{N}\frac{m_n(\dot{\br}^n_t)^2}{2}-V_N+E \right)  \,dt
\end{equation}
with the fixed end points $\bar{\xi}_0$ and $\bar{\xi}_\tau$.  The
connection between equations (\ref{As}) and (\ref{Sf}) suggests that,
for small enough $\tau$,
this functional must be positive definite and have a minimum for some
fixed values of the constants $a$ and $E$.

Now we shall address the question of whether $A_S$ is positive
definite and has a minimum in the context of the physical meaning
of~$E$.  For any function $V_N=V_N(\bar{\br}_t^1,\dots,\bar{\br}_t^N)$
and constant values of $\{m_n\}_{n=1}^{N}$ the Euler-Lagrange
equations for $A_S$ imply that
\begin{equation}\label{kolm13}
\frac{d}{dt}(\sum_{n=1}^{N}\frac{m_n(\dot{\br}^n_t)^2}{2}+V_{N})=0\;,
\end{equation}
which is the well known law of conservation of energy.  Since $E$ is a
constant of motion it may only depend on fundamental constants and
integrals of motion. In our case the energy is generally the only such
integral of motion. Since $V_{N}$ belongs to a very broad class of
functions the only way to ensure that $A_S$ is positive definite is to
require that $E$ contains $V_{N}$ with the plus sign to compensate
$-V_{N}$ in the Lagrangian.  Therefore it is necessary that up to an
insignificant constant, which can be absorbed in (\ref{kolm13}), we
have
\begin{equation}\label{kolm14}
E=\sum_{n=1}^{N}\frac{m_n (\dot{\br}^n_t)^2}{2}+V_{N}\;.
\end{equation}
For sufficiently small $\tau$, this value of $E$ corresponds to a positive definite $A_S$ which has a
minimum as required. Indeed, substitution of (\ref{kolm14}) into
(\ref{As}) gives
\begin{equation}\label{kolm15}
A_S=a\int_0^\tau \sum_{n=1}^{N} m_n (\dot{\br}^n_t)^2\ dt
\end{equation}
which for $a>0$ is a sum of essentially positive terms.  The minimum
should be found using the additional requirement (\ref{kolm14}) as an
auxiliary condition.  To emphasize the analogy with the relativistic
case considered in the next subsection, we reformulate this result in
purely geometrical terms~\cite{Lanczos}. The coordinate
transformation from the Cartesian $(x_n,y_n,z_n)$ to other generalized
coordinates $q_j$ implies that there exists symmetric $m_{jk}$ such
that
\begin{equation}\label{kolm16}
\sum_{n=1}^{N}m_n (\dot{\br}^n_t)^2=
 \frac{\sum_{j,k}m_{jk} dq_j dq_k}{(dt)^2}\;.
\end{equation}  
Using the condition (\ref{kolm14}) we have
\begin{equation}\label{kolm17}
[\sum_{n=1}^{N}m_n (\dot{\br}^n_t)^2dt]^2=
\sum_{j,k}2(E-V_N)m_{jk} dq_j dq_k\;,
\end{equation}
and therefore the symmetric matrix
\begin{equation}\label{kolm18}
 g_{jk}\equiv 2(E-V_N)m_{jk}
\end{equation}
can be used to define a line element of the form
\begin{equation}\label{kolm19}
(dl)^2\equiv\sum_{jk}g_{jk}dq_j dq_k\;.
\end{equation} 
Using the definitions (\ref{kolm18},\ref{kolm19}) and
Eq.~(\ref{kolm17}), we can rewrite (\ref{kolm15}) as
\begin{equation}\label{kolm20}
A_S=a\int_{\xi_0}^{\xi_\tau} dl \;,
\end{equation}
where we emphasize that the sign of $a$ must be chosen to compensate
the sign degeneracy $dl= \pm \sqrt{(dl)^2}$, that is to keep
(\ref{kolm20}) a minimum principle.  Equation (\ref{kolm20}) shows
that the minimum principle is equivalent to the problem of finding a
geodesic path between two fixed end-points $\xi_0$ and $\xi_\tau$ in
the system's configuration space defined by the Riemannian metric
$g_{jk}$.  The metric in its turn was derived using only
\begin{itemize}
\item the {\it SP}$\,$, and
\item the Galilean principle of relativity. \\
\end{itemize}

\subsection{Relativity }

In the previous section we demonstrated our approach using
the example of a conservative nonrelativistic mechanical system.
Our arguments can be summarized into three stages.
First, we used the {\it SP} and the properties of Kolmogorov complexity
to obtain Hamilton's principle of least action
together with the general structure of Lagrangians.
Second, we used the classical arguments by Landau and Lifshitz
to specify the Lagrangians of individual particles. Third,
we constructed a reference computer to check
whether the second stage is consistent with
our choice of Kolmogorov complexity as a measure
of complexity for dynamical laws.

In this section we consider the case of relativistic systems.
The first stage of our arguments is identical to the
one of the nonrelativistic case. This means that
in this section we can start our arguments directly
from the second stage, i.e. consider one particle cases such
as a free relativistic particle and a relativistic particle in an external
gravitational field. To do this this we will need to replace the Galilean
principle with Einstein's principle of relativity. For the third stage 
of the argument we can use the arguments of the previous section
as a template.

We will see that the derivations of this section are
considerably simpler and more natural than in the case of Newtonian
mechanics.  In particular, we do not construct an integrable and twice
differentiable interpolation $L_1$ of the complexity function. More
simplification is achieved in the construction of examples of
reference computers which are consistent with our derivations.
In the
nonrelativistic case we required that the reference computer should be
supplied with a description of the reference frame where the problem
is formulated.  Such a description can be supplied in the form of code
which is appended to the main program. In the nonrelativistic case we
showed that the appended code of fixed length $m\oo{\br}{}_t^2/2+ {\rm
  const}$ does the job. Even though this is a mathematically
consistent requirement, it is rather artificial that the length of the
code depends on the system state. We will see that the analogous
construction in the relativistic case does not require the appended
code at all.  At the end of this section we outline the possibility of
geometrical formulations of our approach and show how one can derive
the theory of particle motion in an external gravitational field.

The physical state of a relativistic mechanical system is described by
the same set of parameters $\xi_k=(\br_k,\dot{\br}_k, t_k)$ as in the
case of a Newtonian mechanical system.  The concept of absolute time,
however, is in deep contradiction with the Einstein principle of
relativity. For this reason it is not convenient to choose time as the
abstract parameter $k$ in Eq.~(\ref{LimEasyLearnable}) as it was done
in the case of Newtonian mechanics. If $c$ is the speed of light, the
Einstein principle of relativity for the case of homogeneous isotropic
space and homogeneous time suggests that the quantity
\begin{equation}\label{kolm21}
(\Delta s)^2=c^2(\Delta t)^2-(\Delta x)^2-(\Delta y)^2-(\Delta z)^2\;,
\end{equation} 
is the same in all inertial reference frames. It is convenient to
choose this quantity for the parameterization in
Eq.~(\ref{LimEasyLearnable}) in the same way as time was chosen in the
case of Newtonian mechanics.  Fixing the limits of parameterization
$s\in [0,\varsigma]$ we have, by analogy with
Eq.~(\ref{IntroduceTau}),
\begin{equation}                        \label{IntroduceSigma}
K_{\bar{\xi}_0}[\Jf]
                =\frac{1}{\varsigma}
                  \sum_{s=0}^{\varsigma-\Delta s}
                  K(\bar{\xi}_{s+\Delta s}|\bar{\xi}_{s})\,\Delta s\;,
\end{equation}
where the sum over $s$ goes from $0$ to $\varsigma-\Delta s$ in steps
of $\Delta s$.

The Einstein principle of relativity requires that the dynamical laws
obtained by minimization of $K_{\bar{\xi}_0}[\Jf]$
must be invariant with respect to
the Lorentz transformations which relate different inertial reference
frames to each other. Considering the case of one free particle, we
also have the requirement of homogeneity of the space and time which
requires that $K(\bar{\xi}_{s+\Delta s}|\bar{\xi}_{s})$ cannot depend
on the time or the coordinates of the particle, i.e. is a function
only of the four-velocity
\begin{equation}
u\equiv (u^0,u^1,u^2,u^3)
\equiv(\frac{\Delta ct}{\Delta s},\frac{\Delta x}{\Delta s},
              \frac{\Delta y}{\Delta s},\frac{\Delta z}{\Delta s})\;.
\end{equation}
The Lorentz transformations can be considered as rotations in
four-dimensional space with the metric $g^{jk}={\rm
  diag}(1,-1,-1,-1)$.  This means that $K(\bar{\xi}_{s+\Delta
  s}|\bar{\xi}_{s})$ cannot depend on the {\it direction} of the
four-velocity. Since $u_ju_kg^{jk}=1$, the absolute value of the
four-velocity is a constant and therefore
\begin{equation}                        \label{OneRelativisticParticle}
K(\bar{\xi}_{s+\Delta s}|\bar{\xi}_{s})
={\rm const} +\frac{\Delta Q}{\Delta s}\;,
\end{equation} 
where $Q$ is an arbitrary function of the system state. This equation
is a relativistic analogue of Eq.~(\ref{OneParticle}). Substitution of
(\ref{OneRelativisticParticle}) into (\ref{IntroduceSigma}) gives
\begin{equation}
K_{\bar{\xi}_0}[\Jf]=a
                  \sum_{s=0}^{\varsigma-\Delta s}\,\Delta s\,.
\end{equation}
Choosing $\Delta s=+\sqrt{(\Delta s)^2}$ we see that $K_{\bar{\xi}_0}$ has a
minimum for negative $a$. In the usual case, when dynamical laws can
be interpolated by twice differentiable functions, the above equation
becomes
\begin{equation}\label{kolm22}
A_S^{\rm rel}=a\int_{0}^{\varsigma}\,ds+{\rm const}\,.
\end{equation}
Minimization of this quantity over possible dynamical laws gives the
known equations of motion for a free relativistic particle.  In other
words, the {\it SP} combined with the Einstein principle of relativity
is enough to obtain the Lagrangian of a free relativistic particle.

As in the case of Eq.~(\ref{OneParticle}), we must show that there
exists a considerable number of reference computers which satisfy
requirement~(\ref{OneRelativisticParticle}).  This can be shown by
repeating the arguments of the previous section. As in the
nonrelativistic case we obtain a family of countably infinitely many
computers which satisfy requirement (\ref{OneRelativisticParticle}).
This time, however, the arguments can be simplified since the
left-hand-side of~(\ref{OneRelativisticParticle}) does not contain
terms quadratic in velocity. Moreover, the constant term in
(\ref{OneRelativisticParticle}) can be absorbed into $\Delta Q/\Delta
s$ which means that the appended code, artificially required in the
Newtonian mechanics, is not necessary in the relativistic case.

It remains to show that $A_S^{\rm rel}$ can be made positive definite.
Writing the constant as $acE\tau$, where $E$ is an integral of motion
and $\tau$ is the time elapsed between the boundary events $\xi_0$ and
$\xi_\varsigma$ we have
\begin{equation} \label{AnyVelocity}
A_S^{\rm rel}=ac\int_0^\tau (\sqrt{1-\dot{\br}^2/c^2}+E)\ dt\;.
\end{equation}

This equation is a relativistic analogue of Eq.~(\ref{As}) in the case
of one particle. As in Newtonian mechanics, it is easy to see that
$A_S^{\rm rel}$ is positive definite when $E$ is equal to the energy
of the system.  Indeed, for small $\dot{\br}^2$ equation
(\ref{AnyVelocity}) becomes
\begin{equation}\label{kolm23}
A_S^{\rm rel}=\int_0^\tau(-mc^2+\frac{m\dot{\br}^2}{2}+...+E) \ dt\;,
\end{equation}
where $m\equiv|a|/c$. For small $\dot{\br}$ the relativistic energy of
a particle is $(mc^2+m\dot{\br}^2/2+...)$; therefore $E$ contains
$+mc^2$ compensating the negative term in (\ref{kolm23}), and
$A_S^{\rm rel}$ is positive as required.  The action remains positive
definite for any values of $\dot{\br}$ because the relativistic energy
grows monotonically with $\dot{\br}^2$.

Looking at equations (\ref{kolm22}) and (\ref{kolm20}), we see that
the problem of identifying a predominant dynamical law is equivalent
to the problem of finding a geodesic path between two fixed end-points
in the system configuration space. The metric of the configuration
space is again determined only by the {\it SP} combined with the
Galilean or Einstein's principles of relativity.  The case of a
particle in an external gravitational field trivially fits this scheme
\cite{LandauF87}. Einstein's principle of equivalence requires that an
external gravitational field can be introduced as an appropriate
change in the metric of space-time, that is as a change in the
expression of $ds$ in terms of $dx$, $dy$, $dz$ and $dt$. Equation
(\ref{kolm20}) has the same form for all such expressions, and the
requirement of minimum of $A_S^{\rm rel}$ gives the standard equations
of motion for a particle in an external gravitational field
\cite{LandauF87}. \\

\section{Discussion } \label{Discussion}

We introduced a new {\it physical} principle -- the Simplicity 
Principle.
It is based on the classical principle of Occam's Razor which is a
cornerstone of the modern theory of induction and machine learning.
Using the Simplicity Principle, we explained the general structure
of the Lagrangian for a composite physical system. In fact, we
explained all generic postulates of the Lagrangian formulation
of physical dynamics. We demonstrated our approach using
the examples of Newtonian mechanics, relativistic mechanics
and the motion of a relativistic particle in an external gravitational field.
We thereby establish a non-trivial link between the Simplicity Principle
and the principle of stationary action.

We have already mentioned that singling out the simplest hypothesis is
not the optimal strategy of inductive inference. Ideally, we should
consider all possible dynamical laws $\{f\}$
weighted in accordance with their complexity $2^{-K_x[f]}$.
This is reminiscent of the Feynman path integrals approach to
quantization.  At present, in quantum field theory a typical
derivation of Feynman path integrals from first principles cannot be
considered mathematically rigorous \cite{Ramond}. There are
problems with convergence and with analytic continuation from
Minkowski space to Euclidean space. Using Kolmogorov complexity
instead of Euclidean action would improve the convergence while
preserving all results that can be attributed to the contributions of
simple laws. Indeed, as suggested by the Kraft inequality, sums
of the type $\sum_f2^{-K_x[f]}$ are convergent: this is not always
the case with Feynman path integrals.  Moreover, there is some
independent evidence~\cite{Woo} that at least a qualitative
relationship between the Euclidean action and Kolmogorov complexity
should exist. Using Kolmogorov complexity instead of the Euclidean
action may also be useful for quantum 
gravity~\cite{Dzhunushaliev} where, among others, the indefiniteness
of the gravitational action is a serious problem~\cite{Gibbons}.
These and other applications of the proposed approach are a
matter for further research. \\

\section*{Appendix } 

To a large extent, the standard derivation of the
Euler-Lagrange equations is based on the well-known
method of integration by parts. 
Here we briefly review its discrete analogue -- Abel's method
of ``summation by parts''. We
then apply this method to derive (\ref{DiscreteNewton}).

\subsection*{Summation by parts}
Let
\begin{equation} \label{DefTtUandTtV}
    \ttU_J\equiv\sum_{k=0}^J \ttu_k
    {\mbox{\ \ and \ }}
    \ttV_J\equiv\sum_{k=0}^J\ttv_k\;,
\end{equation}
where we adopt the usual convention that if $b<a$ then
$\sum_{k=a}^{b}F(k)=0$ for any function $F$.  The integration by parts
method can most easily be demonstrated as a consequence of the
Leibnitz rule
\begin{equation}
d(\ttU\ttV)=\ttU d(\ttV)+\ttV d(\ttU)\;.
\end{equation}
By analogy we therefore compute
\begin{eqnarray}
\ttU_{k}\ttV_{k}-\ttU_{k-1}\ttV_{k-1}
&=&
\ttU_{k}(\ttV_{k}-\ttV_{k-1})
    +\ttV_{k-1}(\ttU_{k}-\ttU_{k-1})\cr
                     &=&\ttU_{k}\ttv_k+\ttV_{k-1}\ttu_k\;.
\end{eqnarray}
Summation from $k=0$ to $k=J$ gives
\begin{equation}                        \label{SumByParts}
\ttU_{J} \ttV_{J} =\sum_{k=0}^{J}\ttU_{k}\ttv_k
+\sum_{k=1}^{J}\ttV_{k-1}\ttu_k\;.
\end{equation}
This ``summation by parts'' formula can be used for manipulating
discrete sums just like the integration by parts is used to manipulate
integrals. In particular, we can derive Eq.~(\ref{DiscreteNewton}) as
a discrete analogue of the Euler-Lagrange equations as follows.  In
Eqs.~(\ref{DefTtUandTtV},\ref{SumByParts}) we replace the abstract
summation index $k$ with the time $t$, as explained for
Eqs.~(\ref{AbsorbTPlusDeltaT}) and~(\ref{IntroduceTau}), to get (using
the earlier defined notation)
\begin{equation}
\ttU_{J}=\sum_{k=0}^{J}\ttu_k
=\sum_{t=0}^{\tau}\ttu_t\equiv\ttU_\tau \;,
\end{equation}
where $\ttu_t=\ttU_t-\ttU_{t-\Delta t}=\Delta \ttU_t$.  Writing
analogous relations for $\ttV_t$ and $\ttv_t$, we can rewrite
Eq.~(\ref{SumByParts}) as
\begin{equation}                        \label{SumByPartsII}
\ttU_{\tau} \ttV_{\tau} =\sum_{t=0}^{\tau}\ttU_{t}\Delta \ttV_t
+\sum_{t=\Delta t}^{\tau}\ttV_{t-\Delta t}\Delta\ttU_t\;.
\end{equation} \\

\subsection*{One particle in a potential }

The case of one particle moving in a potential $V(\bar{\br}_t)$ is
built upon the approximation that the particle interacts with a
massive system which determines the potential, and whose dynamics are
not sensitive to the particle motion. Mathematically, such an approximation
is performed as follows.
The complexity of the dynamical
law $\Jf$ for the whole system has the form
\begin{equation}                        \label{sum}
K_{\bar{\xi}_0}[\Jf]\propto \sum_{t=0}^{\tau -\Delta t}
K(\ooo{\br}_{t+\Delta t},
\ooo{\br}{}^1_{t+\Delta t},\dots \ooo{\br}{}^N_{t+\Delta t}
|\bar{\xi}_t)\Delta t\;,
\end{equation}
where the upper indices from 1 to $N$ refer to the particles of the massive
system.
Formally separating our particle from the massive system we
have from~(\ref{Sf})
\begin{equation}
K(\ooo{\br}_{t+\Delta t},
\ooo{\br}{}^1_{t+\Delta t},\dots \ooo{\br}{}^N_{t+\Delta t}
|\bar{\xi}_t)
=\frac{m\,(\oo{\br}_t)^2}{2}
            -\Big[V_{N+1}-\sum_{n=1}^{N} \frac{m_n\,(\oo{\br}{}^n_t)^2}{2}\Big]
                                       +{\rm const}\;.
\end{equation}
To make the above approximation we assume that
the equations of motion for the massive system are known
and are not affected by the motion of our particle.
This means that variables $\{ \oo{\br}{}^n_t \}$
can be eliminated and the expression in square brackets
can be replaced by an {\it effective} interaction $V(\bar{\br}_t)$:
\begin{equation}
K(\ooo{\br}_{t+\Delta t},
\ooo{\br}{}^1_{t+\Delta t},\dots \ooo{\br}{}^N_{t+\Delta t}
|\bar{\xi}_t)
\approx
K(\ooo{\br}_{t+\Delta t}|\bar{\br}_t,\oo{\br}_t,t)
\equiv\frac{m\;(\oo{\br}_t)^2}{2}
                   -V(\bar{\br}_t)+{\rm const}\;.
\end{equation}
Although useful in practice,
this approximation ruins the connection between the interaction
and the mutual information. Effective interaction $V(\bar{\br}_t)$
has contributions from the kinetic energy terms and it
strongly depends on the assumed equations of motion.
  Thus,
the information-theoretic interpretation of the interaction terms in
the Lagrangian is only valid for the fundamental interactions:
before any approximations are made.

For simplicity, we shall consider the case of one dimensional motion,
where $\oo{\br}_t$ and $\bar{\br}_t$ can be considered as scalars.
Generalization to the multidimensional case is essentially trivial.
We define the discrete variation in absolute analogy with standard
variational calculus
\begin{equation}
\delta V(\bar{\br}_t) \equiv 
V(\bar{\br}_t+\delta\bar{\br}_t)-V(\bar{\br}_t)\;,
\end{equation}
where $\delta\bar{\br}_t$ is a virtual change of the function
$\bar{\br}_t$.  For instance if $T(\oo{\br}_t)\equiv
m\,(\oo{\br}{}_t)^2/2$ then
\begin{equation}                        \label{deltaT}
\delta T(\oo{\br}_t)=\frac{m}{2}
[(\oo{\br}_t+\delta\oo{\br}_t)^2-(\oo{\br}{}_t)^2]
=m \oo{\br}_t \delta\oo{\br}_t+ { m\,(\delta\oo{\br}{}_t)^2 / 2}\;.
\end{equation}
To minimize the sum (\ref{sum}) we require that, up to second order in
$\delta\bar{\br}_t$ and $\delta\oo{\br}_t$,
\begin{equation}                        \label{NoTauTerm}
\delta \sum_{t=0}^{\tau -\Delta t}
\left[ T(\oo{\br}_t)- V(\bar{\br}_t)\right]\Delta t
\equiv \sum_{t=0}^{\tau -\Delta t}
 \left[ \delta T(\oo{\br}_t) - \delta V(\bar{\br}_t)\right]\Delta t=0\;.
\end{equation}
Because we do not vary the functions $\bar{\br}_t$ and $\oo{\br}_t$ at
the end points $t=0$ and $t=\tau$, we have $\delta T(\oo{\br}_\tau)=0$
and therefore (\ref{NoTauTerm}) is equivalent to
\begin{equation}                        \label{AddTauTerm}
\sum_{t=0}^{\tau}
 \delta T(\oo{\br}_t) \Delta t
 - \sum_{t=\Delta t}^{\tau}
     \delta V(\bar{\br}_{t-\Delta t}) \Delta t=0\;.
\end{equation}
Noticing that
\begin{equation}
\delta \oo{\br}_t
 =\frac{\delta\bar{\br}_t-\delta\bar{\br}_{t-\Delta t} }{\Delta t}
 =\frac{\Delta \delta \bar{\br}_t }{ \Delta t}
\end{equation}
we have
\begin{equation}
\sum_{t=0}^{\tau}\delta T(\oo{\br}_t)\,\Delta t
= \sum_{t=0}^{\tau}
   \frac{\delta T(\oo{\br}_t) }{ \delta \oo{\br}_t }
\,\Delta\delta\bar{\br}_t\;,
\end{equation}
where it is understood that $\delta T(\oo{\br}_\tau) / \delta
\oo{\br}_\tau =0$ as required by the transition from
Eq.~(\ref{NoTauTerm}) to Eq.~(\ref{AddTauTerm}).  Setting $\ttU_t=
\delta T(\oo{\br}_t) / \delta \oo{\br}_t $ and $\ttV_t=\delta
\bar{\br}_t$ for all $t$, we use summation by parts
(\ref{SumByPartsII}) to show
\begin{equation}                        \label{ApplySumByParts}
\sum_{t=0}^{\tau}
   \frac{\delta T(\oo{\br}_t) }{ \delta \oo{\br}_t }\,\Delta\delta\bar{\br}_t
=-\sum_{t=\Delta t}^{\tau} \delta\bar{\br}_{t-\Delta t}\,
 \Delta \frac{\delta T(\oo{\br}_t) }{ \delta \oo{\br}_t }\;.
\end{equation}
Combining Eqs.~(\ref{AddTauTerm}) and (\ref{ApplySumByParts}), we
have, up to second order in $\delta\bar{\br}_t$,
\begin{equation}
\sum_{t=\Delta t}^{\tau}
\left[ \frac{\Delta}{\Delta t}
 \frac{\delta T(\oo{\br}_t) }{ \delta \oo{\br}_t } 
+\frac{ \delta V(\bar{\br}_{t-\Delta t}) }{ \delta \bar{\br}_{t-\Delta t} }
  \right]\delta\bar{\br}_{t-\Delta t}\, \Delta t=0\;.
\end{equation}
This must be true for arbitrary values of $\delta\bar{\br}_{t-\Delta
  t}$ and therefore we demand
\begin{equation}
 \frac{\Delta}{\Delta t}
 \frac{\delta T(\oo{\br}_t) }{ \delta \oo{\br}_t } 
+\frac{ \delta V(\bar{\br}_{t-\Delta t}) }{ \delta \bar{\br}_{t-\Delta t} }=0\;.
\end{equation}
Now substituting $T(\oo{\br}_t)= m\,(\oo{\br}{}_t)^2/2$ and using
Eq.~(\ref{deltaT}) we have (\ref{DiscreteNewton}) as required. \\

\section*{Acknowledgments}
I am very grateful to Dominik Janzing, R.~E.~Wilson,
James~B.~Hartle, A.~S.~Johnson, Todd A. Brun
 and especially to Jens G. Jensen and
R\"udiger~Schack for their interest, support and discussions.  A part
of this work was completed at the Isaac Newton Institute in Cambridge,
the hospitality of which is gratefully acknowledged.

\end{document}